\documentstyle[aasms4,epsfig,epsf]{article}
\begin{document}


\title{Constraints on the Delayed Transition to Detonation in Type Ia Supernovae}
\author{A.M. Lisewski, W. Hillebrandt}
\affil{Max\--Planck\--Institut f\"ur Astrophysik,
Karl-Schwarzschild\--Str. 1, 85740 Garching, Germany}
\author{S.E. Woosley}
\affil{UCO/Lick Observatory, University of California Santa Cruz,
Santa Cruz, CA 95064, USA}


\begin{abstract}
We investigate the possibility of a delayed detonation 
in a type Ia supernova under the assumption that  the transition to detonation
is triggered by turbulence only. Our discussion is  based
on the Zeldovich mechanism and suggests that typical
turbulent velocities present during the explosion are not strong enough to allow
this transition to occur. Although we are able to show that in carbon-rich matter (e.g., $X(^{12}$C$) =
0.75$) the possibility of a deflagration to detonation
transition (DDT) is enhanced, even in this case the turbulent 
velocities needed are larger than the expected value of $u'(L) \approx 10^7 \mbox{cm
s}^{-1}$ on a length-scale of $L \approx 10^6$ cm.
Thus we conclude that a DDT may not be a common event during a
thermonuclear explosion of a Chandrasekhar-mass white dwarf.
\end{abstract}


\section{Introduction}

The question  whether a detonation occurs during the thermonuclear explosion of a white dwarf 
of Chandrasekhar mass has already been
discussed extensively. 
Many existing models assume that such a transition does
occur in certain situations during the explosion. Some of
these models, the so-called delayed-detonation models,
assume that at a critical value of a model parameter
(usually the density) a previously subsonic burning front turns
into a detonation. In comparison with observables, like lightcurve
shape, peak brightness or element abundances, the delayed-detonation 
scenario seemed to comply well (e.g., \cite{hoef97}). However, a conclusive answer
whether this transition is physically possible has not been given yet. 

In a thermonuclear supernova, the explosion energy is produced by a moving
thermonuclear flame that burns carbon and oxygen into higher mass elements.
According to our present understanding at least at high densities, i.e. in the inner core
($ \rho > 10^8 \mbox{g cm}^{-3}$) of the white dwarf, the burning mode 
of this flame is a deflagration, where the burning front can locally
be described as a laminar flame. The properties of laminar flames
in white dwarf matter are well known (\cite{timmwoos92}). One general feature is that with lower
density the speed of a laminar flame decreases rapidly while at the same 
time its thickness increases. On the other hand, numerical models of 
type Ia supernovae taylored to reproduce lightcurves and spectra require a flame velocity that is nearly independent
of the density in the unburned material. 
Thus there must be a mechanism  that leads to an effective
flame velocity being independent of its laminar properties.
It is commonly believed that turbulence provides
this mechanism. However, quantitative predictions of the turbulent
flame velocity are not easy to obtain because they depend
on many (often very uncertain) parameters such as density, nuclear
compositions, turbulent properties, etc. It is not
even clear if the flame speed remains always subsonic
or if it might turn into a detonation. Leaving out scenarios
like the prompt initial detonation (see, for example, \cite{imsh99}
and references therein) and the pulsating delayed detonation (e.g., \cite{khoketal97}), we focus on the problem
whether turbulence alone is able to trigger a supersonic burning
wave accompanied by a hydrodynamic shock.

In order to achieve this goal, we introduce a method to represent the structure
of a flame moving through a turbulent medium. 
For a given physical state
of the unburned matter, and for a specified turbulent
energy, the model adopts the most 
optimistic assumptions for the transition from a deflagration to a detonation.
We then investigate if such a transition happens under these special
prerequisites.
Cases where no transition is observed therefore provide
quantitative limits on some relevant
physical parameters, like nuclear composition or turbulence intensity.
These limits constrain the regime where a detonation triggered by
turbulence might occur.

This letter is structured as follows: First we briefly recall some
properties of carbon-flames at densities around $ 10^7 \mbox{g
cm}^{-3}$ where according to our present understanding (see,
e.g. \cite{khoketal97} and \cite{niemwoo97}) the chances of a DDT are
the highest. Then we introduce our method to
represent turbulence effects on subsonic flames. In this way we obtain some
turbulent flame profiles which we eventually implement as initial conditions
in a gas-dynamic system to study their temporal evolution.
Finally, we discuss some implications of our results.

\section{The model}

The structure of a flame moving through a turbulent medium crucially
depends on the characteristic turbulent lengthscales, like the scale where
the biggest turbulent motions are produced and the dissipation scale
where
turbulent motion is smeared out by microscopic diffusion, as
well as on the amount of turbulent kinetic energy that drives the turbulent motion. 
Of course, in case of turbulence driven by inertia these scales  become
the inertial 
scale $ L$ and the Kolmogorov scale $ l_{\rm k}$. On the other hand,
it is important to know how these typical turbulent entities
relate to the given chemical and thermodynamical properties that are 
responsible for the burning process itself.\\
Here we set up a model that represents this relation in a certain 
way. Consider a flame -- laminar or turbulent -- in the incompressible
case, that is a flame with no pressure jump across it. 
Then the burned and unburned
states of matter are given by the density, temperature and
nuclear composition of the unburned state and by the available amount of
energy released by nuclear reactions under these conditions.
The only difference between a turbulent and a laminar flame is 
the spatial and temporal distribution of density, temperature and
nuclear abundances within the interface between fuel and ashes, i.e.
the flame profile.
For instance, in a laminar flame temperature is a monotonically decreasing
function in the direction of the unburned material, whereas in the turbulent
case fluctuations can cause this function to have sharp spikes and
dips. 

Turbulence has two main impacts on the burning process.
First, there is an enlargement of
the flame surface caused by turbulent stretching and wrinkling
leading to a higher effective flame speed. Second, in cases
of higher turbulence intensity, vortices can even mix burned or burning 
material with unburned matter. This happens when they are fast enough to 
carry reactive material away from the flame before
this material is burned completely. As a direct consequence one observes a 
broadening of the flame profile. Now the flame structure is
superposed by fluctuations coming from 
turbulent motion. However, in this work we neglect a direct 
representation of fluctuations and instead concentrate 
on one net effect of the latter, namely the 
enhanced heat and mass transport that actually causes the
broadening of the flame. The resulting spatial structure
of the flame and in particular its size play a crucial role
in the DDT problem.

The profile of an incompressible laminar flame is described by the set of 
functions $ T_{\rm l}(x;t)$, $\rho_{\rm l}(x;t)$, $Y_{i,{\rm l}}(x;t)$,
 which give the spatio-temporal
distributions of temperature, density and nuclear number density of the $i$th 
species at a constant pressure $ P$.
These functions are solutions of the following system of equations:
\begin{equation}
\label{abund}
\frac{\partial Y_{i,{\rm l}}}{\partial t} = \sum_{j,k} - Y_{i,{\rm l}} Y_{k,{\rm l}} \lambda_{jk}(i) 
				+ Y_{i,{\rm l}} Y_{k,{\rm l}}
				\lambda_{kj}(i) \,,
\end{equation}
\begin{equation}
\label{energ}
\frac{{\partial} T_{\rm l}}{\partial t} = \frac{1}{\rho c_{\rm p}}\frac{\partial
	}{\partial x} \left( \sigma \frac{\partial T_{\rm l}}{\partial x}
	\right) - P \frac{\partial}{\partial t}\frac{1}{\rho}
					+ \frac{1}{c_{\rm p}}{\dot{S}} \,,
\end{equation}
\begin{equation}
\label{enuc}
\dot{S} = N_A \sum_{i} \frac{{\rm d}Y_{i,{\rm l}}}{{\rm d}t} B_i \,.
\end{equation}
Herein, $ c_{\rm p} = c_{\rm p}(T_{\rm l},\rho_{\rm l},Y_{i,{\rm l}})$ is the heat capacity, $
\sigma = \sigma(T_{\rm l},\rho_{\rm l},Y_{i,{\rm l}})$ 
the thermal conductivity, $ B_i$ the binding energy and 
$ Y_{i,{\rm l}}$ is the number density
of the $ i$th nucleus while $\lambda_{ij}$ denotes the 
the reaction rate of nuclei $ i$ and $ j$.
Our reaction network consists of seven species, viz. $^4$He, $^{12}$C,
$^{16}$O, $^{20}$Ne, $^{24}$Mg, $^{28}$Si, and $^{56}$Ni.
 
The direction of the flame is chosen to be such that the completely burned
state is at $ x = -\infty$.
In a steady state the flame 
represented by these equations propagates with a constant velocity $ s_{\rm l}$.
Thus in a co-moving frame the functions $ T_{\rm l}$, $ \rho_{\rm l}$ and 
$ X_{i,{\rm l}}$ are constant in time and it is convenient
to place the origin of the co-moving coordinate system at the point of
maximum energy generation.

\begin{figure*}[ht]
\figurenum{1}
\label{tindeps}
\epsscale{.8}
\plotone{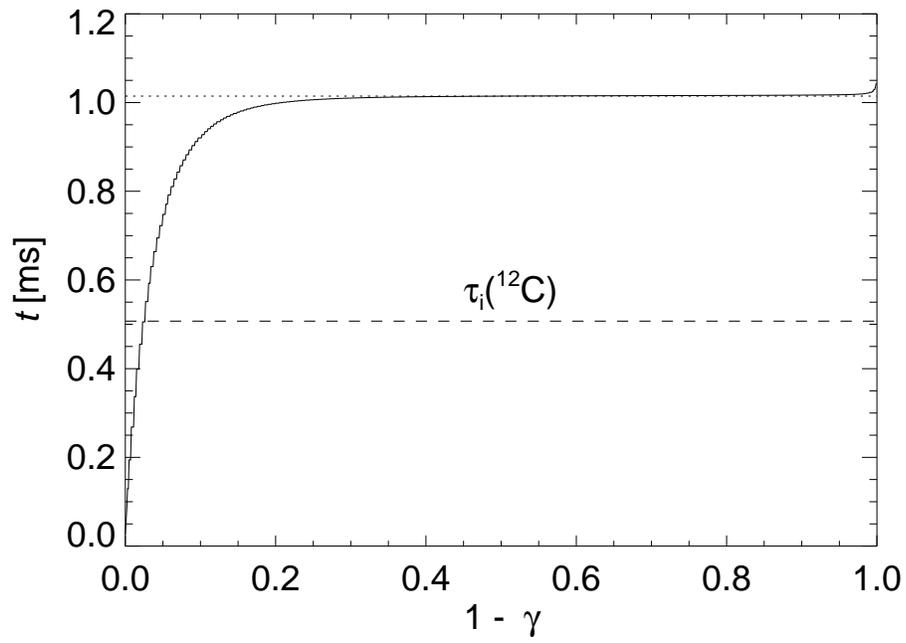}
\caption{The estimated time for burning a fraction $ 1 - \gamma$ 
of an initial carbon abundance $ X(^{12}\mbox{C}) = 1/2$ at
an initial temperature of $ 1.7 \times 10^9$ K and a density of
$ 2.3 \times 10^7 \mbox{g cm}^{-3}$. The resulting value of the
induction time, $ \tau_i$, is shown by the dashed line.}
\end{figure*}

The flame profile defines also a spatial distribution of 
the nuclear reaction timescale of the $i$th element, $ \tau_{\rm m}(i)$, 
by
\begin{equation}
\label{tnucm}
\tau_{\rm m}(i) = \left(\frac{Y_i}{\dot{ Y_i}}\right)_{T, \rho} \,, 
\end{equation}
where the time derivative is taken at constant temperature and density. 
In this work we are particulary interested in C+O matter at densities
around $ 10^7 \mbox{g cm}^{-3}$, where by far the most relevant nuclear reaction
is $ ^{12}\mbox{C} + \,^{12}\mbox{C}$. Burning of heavier elements takes
place on timescales much longer than $ \tau_{{\rm m}}(^{12}{\rm C})$
and releases much less energy. 
We will therefore always refer to carbon-burning 
in the following.\\
Note that the equation (\ref{tnucm}) oversestimates the nuclear
burning time, because burning a small fraction of the given carbon already significantly raises
the temperature, which in turn increases the reaction rate.
Equation (\ref{tnucm}) assumes that temperature remains 
constant during the reaction, neglecting any self-accelerating effects
of nuclear burning. Thus to obtain a more realistic estimate on the
nuclear burning time we will refer to the so-called induction time of
nuclear burning. It is known (e.g., \cite{khok91}) that the process
of explosive nuclear burning can be subdivided into two temporal
stages. During the first phase after ignition, at times shorter
than the induction time $ \tau_{\rm i}$, the temperature increases
slowly (linearly) in time, and the corresponding energy production 
is relatively small.\\
After this, during the explosive phase, energy production evolves
quasi-exponentially in time. Thus only the induction time is actually
needed to burn  a fluid element completely, because $ \tau_{\rm i}$ is
nearly equal to the time it takes to burn most of the  fuel. Self-acceleration of nuclear burning leads
to the relation $ \tau_{\rm i} \ll 
\tau_{\rm m}$. The ratio of the induction
and the constant temperature timescale can approximately  be given  by an 
analytic expression known as the Frank-Kamenetskii factor. However, for 
our purposes we calculate $ \tau_{\rm i}$ numerically
using the equation
\begin{equation}
\label{taucm}
\int\limits_{0}^{2 \tau_{\rm i}} \,{\dot{Y}_{^{12}{\rm C}}}\,{\rm d}t = \gamma\, Y_{^{12}{\rm C}}(t = 0) \,,
\end{equation}
where $ \gamma$ denotes the total fraction of burned material.
Based on equation (\ref{taucm}), Figure \ref{tindeps} represents
the time  required to burn a fraction
$(1 - \gamma)$ of the initial fuel amount and it clearly shows 
that for any value $ \gamma \in [0,2,0.9]$ this time turns out to be
alomst the same. 
We therefore use equation (\ref{taucm}) -- with an arbitrary  
choice of $ \gamma \in [0,2,0.9]$ -- to define the induction time 
$ \tau_{{\rm i},^{12}{\rm C}}$
as {\it half} of the time it takes to burn most of the fuel.
This definition ensures that for times smaller than $ \tau_i$
the burning process evolves almost linearly.
 
For a phenomenological description of turbulence we 
 assume that the turbulent velocities obey Kolmogorov's scaling law.
In particular, this implies the following scaling
of the r.m.s. turbulent velocity fluctuations $ u'(l)$:
\begin{equation}
u'(l) = u'(L) \,\left(\frac{l}{L}\right)^{1/3}\,,
\end{equation}
with $  l_{\rm k} = L \,\mbox{Re}^{-3/4} < l < L$, where $ L$ is the integral
length-scale and Re is the turbulent Reynolds number. 
In order to model the effect of turbulent motions on a flame
front we make two additional assumptions:

1. Turbulent vortices are viewed as a transport mechanism
of heat and mass. Thus a fluid element can be carried by a vortex 
over a distance $ l$ after a time $ \tau_l = 1/2 \,(l/u'(l))$, which
is half of the eddy's turn-over time. In case of a one-dimensional
and co-moving representation of a flame, this mechanism transports 
heat and nuclear species located at $ x > 0$ in the positive
direction, while fluid elements placed at $ x < 0$ are transported
in the negative one.

2. Since the spatial distribution of the induction time  $ \tau_{\rm
i}(x)$ is a single-valued function in the regions $ x > 0$ and $ x < 0$,
every fluid element has its unique value of $ \tau_{\rm i}(x)$.
Then the distance $ l'_\tau$ over which a fluid element located 
at $ x$ can be moved is determined by the equation $ \tau_l = \tau_{\rm i}$
leading to 
\begin{equation}
\label{led1}
l'_{\rm \tau}(x) = \frac{\left(2 \, u'(L) \,\tau_{\rm i}\right)^{3/2}}{L^{1/2}} \,. 
\end{equation}
In this approach  $ l'_{\rm \tau}(x)$ is not allowed to exceed the
integral length, 
$ L$, neither should it get smaller than the Kolmogorov length, $
l_{\rm k}$
\footnote{
The reason why we do not consider length-scales larger than $ L$ is
that we actually do not know how the turbulent spectrum of velocity
fluctuations looks like at long wavelengths. Rayleigh-Taylor unstable
structures (bubbles) in the non-linear stage of this instability
may cause additional effects on the turbulent behavior (see, e.g., 
\cite{niemwoo97}) leading to significant deviations from isotropy and
homogeneity.}.
Furthermore, because  the flame itself
moves on during the time $ \,\tau_{\rm i}$ (at least with its
laminar velocity $ s_{\rm l}$), we have 
to subtract the distance $ s_{\rm l} \tau_{\rm i}$ from the original value
of $ l'_{\rm \tau}(x)$. This results in the expression
\begin{equation}
\label{led2}
l_{\rm \tau}(x) = H(\chi_{[l_{\rm k}, L]}(l'_{\rm \tau}) \,l'_{\rm
\tau}\,  - s_{\rm l}\tau_{\rm i})\!
\left[ \chi_{[l_{\rm k}, L]}(l'_{\rm \tau}) \,l'_{\rm \tau}\,  - s_{\rm l} 
\,\tau_{\rm i} \right]\,, 
\end{equation}
where $ \chi_{[l_{\rm k}, L]}$ is the characteristic function of the interval
$[l_{\rm k}, L]$ (This is, $ \chi_{[l_{\rm k}, L]}(x) = 1 \,\mbox{for}\, x \in[l_{\rm k},
L]$, and 0 otherwise.) for giving a cut-off for scales smaller
than $ l_{\rm k}$ and larger than $L$. $ H$ is Heaviside's function representing the
obvious fact that negative values of $ l_{\rm \tau}(x)$ are not
admissable.

Using equation (\ref{led2}), the turbulent flame profile is 
given by
\begin{equation}
        \label{tprofil} \left. \begin{array}{cc}
\{\rho_{\rm t},T_{\rm t},Y_{\rm t}\}(x + l_{\rm \tau}(x)) & x > 0\\
\{\rho_{\rm t},T_{\rm t},Y_{\rm t}\}(x - l_{\rm \tau}(x))  & x < 0 
				                           \end{array}
                        \right\}
= \{\rho_{\rm l},T_{\rm l},Y_{\rm l}\}(x)
\end{equation}

Our first assumption was that there is additional heat/mass-transport
due to turbulent eddies leading to a broadening of the
flame. But it ignores the fact that turbulence is actually isotropic,
which means that eddies occur in any direction without a preferred one. 
However, as long as the flame's reaction zone
is relatively well localized there will always be a flame surface defined 
by the points of maximum energy production. 
Thus there is a preferred direction locally defined by
the normal vector of the flame surface.
Turbulence randomly 
drags and shifts this surface but in a locally co-moving frame turbulent motion
increases the thermal and nuclear transport in directions normal
to it. Heuristically, this can be seen by introducing a
local {\it eddy-diffusivity} $ D_{\rm t}(x) = u'(x) x$, where $ x$ 
denotes the distance from the flame surface. Thus turbulence imposes
an enhanced heat flux ahead of the flame's surface. 
In the reaction region the induction times are so short that
turbulent mixing hardly changes its structure. Well behind the thin 
reaction zone almost
all of the carbon is already burned. Thus turbulent motions in
regions behind this zone actually stirrs pure ashes.
In the context of the present work it is therefore  sufficient
to consider effects of turbulence only for $ x > 0$.

One may argue that in reality there is not just one flame profile 
along the normal direction of a flame surface, but there is a random 
superposition of flames,
fuels and ashes, giving rise to strong fluctuations in temperature
and nuclear abundances. However, as was stated before,
we do not intend to give a full description of the structure of a
turbulent flame. It is our aim to model a situation which is most favourable
for carbon detonation. In order to trigger 
a detonation based on the Zeldovich mechanism (\cite{zel70}) one needs 
a region of a certain
critical size with a rather uniform  temperature and fuel fraction. 
To be more specific, a non-uniform spatial region
of induction times must be present such that within
this region spontaneous burning reaches a critical velocity threshold.
Let $\sigma(\tau_i)$ be the resulting variance of the induction
times. Then this region must be at least of size $ \sigma(\tau_i) \,a_{\rm s}$,
where $ a_{\rm s}$ is the local sound velocity (see, \cite{khok91}).
In the presence of temperature and fuel 
fluctuations a successful formation of such a region is strongly suppressed,
because the rate and the amplitude of fluctuations grow with scale.
Therefore one expects that such homogeneous regions are already
torn apart into smaller sections before they can ever reach the
critical size (\cite{niem99}).
Our model aims at describing the unlikely situation that
such strong fluctuations are not present and therefore we could find
detonations  which in reality would not happen.
On the other hand, if certain physical conditions described by a set of parameters
$ \rho$, $s_{\rm l}, \mbox{Re}, L, l_{\rm k}$, etc., make a
successful detonation impossible in our model, it indicates
that under the same prerequisites a detonation in a real turbulent
flame would not happen either. 

In order to determine a relation between turbulence intensity 
and the resulting shape of the flame we make one more assumption. 
The main motivation 
for it is the desire to relate an undisturbed (laminar) flame profile
with a time setting a maximum limit for the turbulent transport.
In fact, the maximum time for which a fluid element with  a significant amount of fuel can be transported 
by turbulent motion is $ \tau_{\rm i}$, i.e. its induction time. 
During this period
 it covers a distance $ l_{\rm \tau}$ given by the size of a certain 
eddy, c.f. equation (\ref{led1})
\footnote{It should be remarked
that $ l_{\tau}$, given by equation (\ref{led1}),
is the maximum distance for turbulent transport at a given
$ \tau$. That means that bigger eddies are not able to carry
a fluid element further out than $ l_{\tau}$. This can be
easily verified when Kolmogorov scaling is assumed.}
. Considering  this for every
fluid element within a laminar flame, one can derive
a formula for the shape of the turbulent flame, which is exactly
equation (\ref{tprofil}). This equation already involves  a
simplification because we use the induction 
time in order to estimate $ l_{\rm \tau}$. As mentioned above,
the thermodynamic state of a fluid element
changes only  slightly after a period of time given by
$ \tau_{\rm i}$, i.e. $ T_{\rm l}(t+ \tau_{\rm i}) \approx T_{\rm l}(t)$ and
$ Y_{^{12}{\rm C},\rm{l}}(t + \tau_{\rm i}) \approx Y_{^{12}{\rm
C},{\rm l}}(t)$.
Thus the state of the fluid element after being transported over a 
distance $l_{\rm \tau}$ is nearly the same as its
original state. However, taking into account significantly longer
times than $ \tau_{\rm i}$ would lead to strong spatial fluctuations in
temperature and composition.
For instance, assume that a fluid element is transported over
a distance $ l_{\rm \tau}$ where $ \tau \approx 2 \tau_{\rm i}$.
Furthermore consider a small fluctuation in time 
$ \delta\tau \ll \tau$. Then we have
$l_{\tau \pm\delta\tau} \approx l_{\tau}$ but $ T(\tau \pm \delta\tau)$ 
might be much bigger or smaller then $ T(\tau)$ depending on the sign
of the fluctuation. Of course, spatial fluctuations in carbon
composition would be even stronger in amplitude.
Again, based on the same arguments as given above, these 
fluctuations would make the formation of a detonation much more unlikely.
Therefore for the purpose of this letter it is sufficient to represent
flame-turbulence interaction in
the linear regime characterized by the induction time scale. 

\begin{figure*}[ht]
\figurenum{2}
\label{preh}
\plotone{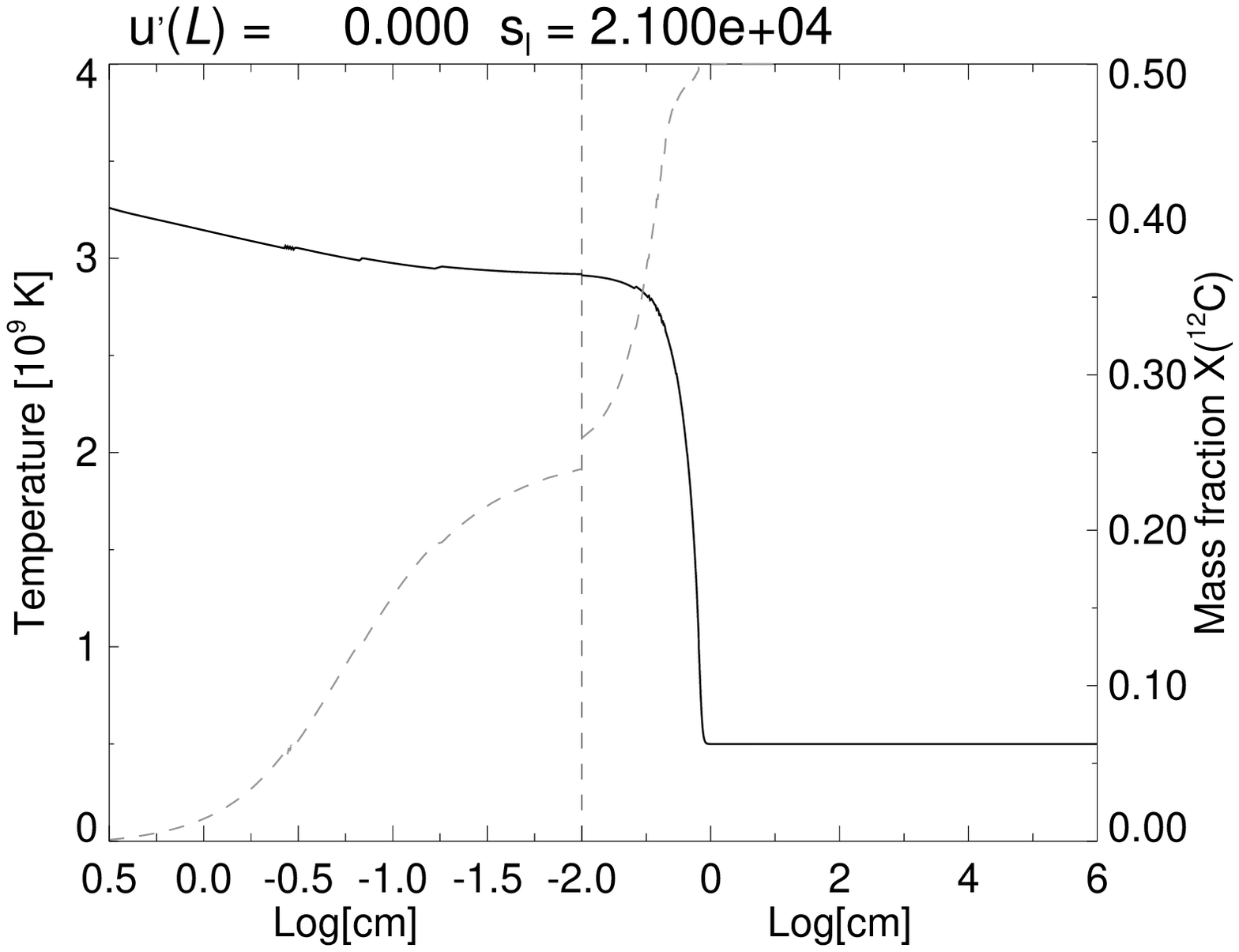}\\
\plotone{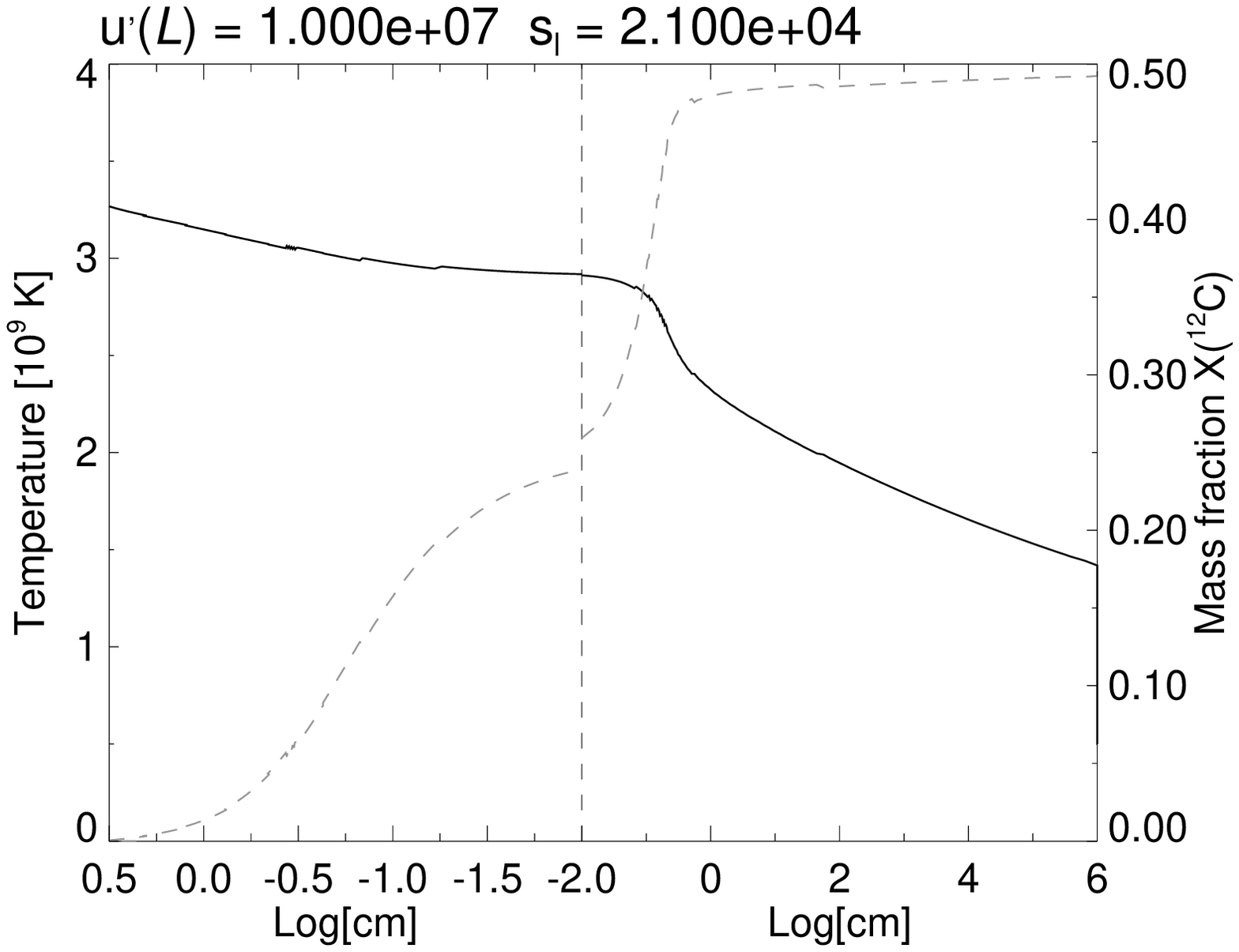}
\caption{Estimated shapes of turbulent flames. A laminar flame profile
(upper panel) and a turbulent flame profile (lower panel). Both
represent a flame at a fuel density of $ 2.3 \times  10^7 
\,\mbox{g cm}^{-3}$.}
\end{figure*}

Two of the resulting turbulent profiles are shown in Figure
\ref{preh}. One sees that the innermost region of the flame is hardly
disturbed by turbulent motions even at turbulent velocities of
$ u'(L) = 10^7 \mbox{cm s}^{-1}$. In contrast, regions
of slow reactions are widely extended causing a huge preheated
zone ahead of the flame.

\section{Testing for Detonations}

In this section we use the results stated above to implement
them as initial conditions for the solution of the fully compressible hydrodynamical
equations. The question then is under which conditions the
flame evolves into a detonation. Similar investigations have
already been  done (\cite{khok91}, \cite{niemwoo97}). In these studies
 the initial conditions were
parametrized by the width of a non-uniformly burning region. It 
was shown that there exists -- for given density and fuel composition --
a minimum (critical) length scale such that non-uniform burning on scales
larger than this leads to a detonation. However, in  these investigations
the important question of how  such a region
might form  was not addressed. Only weak necessary conditions limiting the
strength of the turbulence-flame-interaction have been given. For instance,
the authors demand a {\it breaking up} of the flame
(\cite{khoketal97}) or burning in the 
{\it distributed regime} (\cite{niemwoo97}).\\
Our model establishes a direct relation between turbulence
intensity and the shape of the flame profile. Using this relation
we are able to give necessary conditions for a transition to detonation.

The system of conservation laws that we solve numerically is
\begin{eqnarray}
\frac{\partial \rho}{\partial t} + \nabla \cdot (\rho U) = 0 \,,\\
\frac{\partial \rho U}{\partial t} + \nabla \cdot (\rho U U) + \nabla P = 0 \,,\\
\frac{\partial E}{\partial t} + \nabla \cdot [(E + P) U ] = 0 \,,\\
\frac{\partial Y_i}{\partial t} + (U \cdot \nabla) Y_i +
				\sum_{j,k}  Y_i Y_k \lambda_{jk}(i) 
				- Y_i Y_k
				\lambda_{kj}(i) = 0 \,,
\end{eqnarray}
where $ U$ is the fluid velocity and $ E = \epsilon + \rho U^2/2 +
\rho N_A \sum B_
i Y_i$ is the total energy density defined as the sum of the internal, 
the kinetic and the binding energy term. These equations are coupled
to the same equation of state and nuclear reaction network as already
used in the reactive diffusion equations (\ref{abund}) and
(\ref{energ}). They are solved
in a one-dimensional planar geometry with outflow boundary conditions as
well as in spherical geometry. In the latter case we use a reflecting
inner boundary and an outflow condition on the outside.
As a numerical scheme we used PROMETHEUS (\cite{fryx89}), a second order explicit in 
time solver for reactive hydrodynamics.\\
We consider three different densities, namely  $ 2.3, 1.3$ and $ 0.8
\times 10^7\mbox{g cm}^{-3}$, and three different compositions of the
fuel: $
X(^{12}{\rm C})/X(^{16}{\rm 
O}) = 1,1/3$ and $3$. The turbulent velocity fluctuations in an exploding Chandrasekhar-mass white
dwarf are  (see, \cite{niemwoo97}, \cite{khoketal97}) believed to be
$ u'(L) \approx 10^6 - 10^7 \mbox{cm s}^{-1}$ on an integral lengthscale
of $ L \approx 10^6 \mbox{cm}$. In all our numerical computations
we keep the integral scale fixed at this particular value and the only 
turbulent variable remains the amplitude turbulent velocity fluctuations $ u'(L)$.\\

Given initial conditions as described above we need to find a natural time
limit for evolving the hydrodynamics. Since
the flame dynamics is embedded into the global dynamics of the 
star, we adopt this limit to be the 
hydrodynamical timescale, $\tau_{\rm h}$, of the exploding white
dwarf. After this time has elapsed the stellar density has dropped significantly and  
our originally assumed value of the density in the unburned material
is no longer valid. Thus if no detonation is observed after an elapsed time
$ \tau_{\rm h}$ it is fair to conclude that for the  conditions investigated
no detonation can occur.
We use a value of $ \tau_{\rm h} = 0.02$ s which is 
consistent with estimates from direct numerical calculations of  
exploding Chandrasekhar-mass white dwarfs (\cite{rei99}).

\begin{figure*}
\figurenum{3}
\label{noddt2350}
\plotone{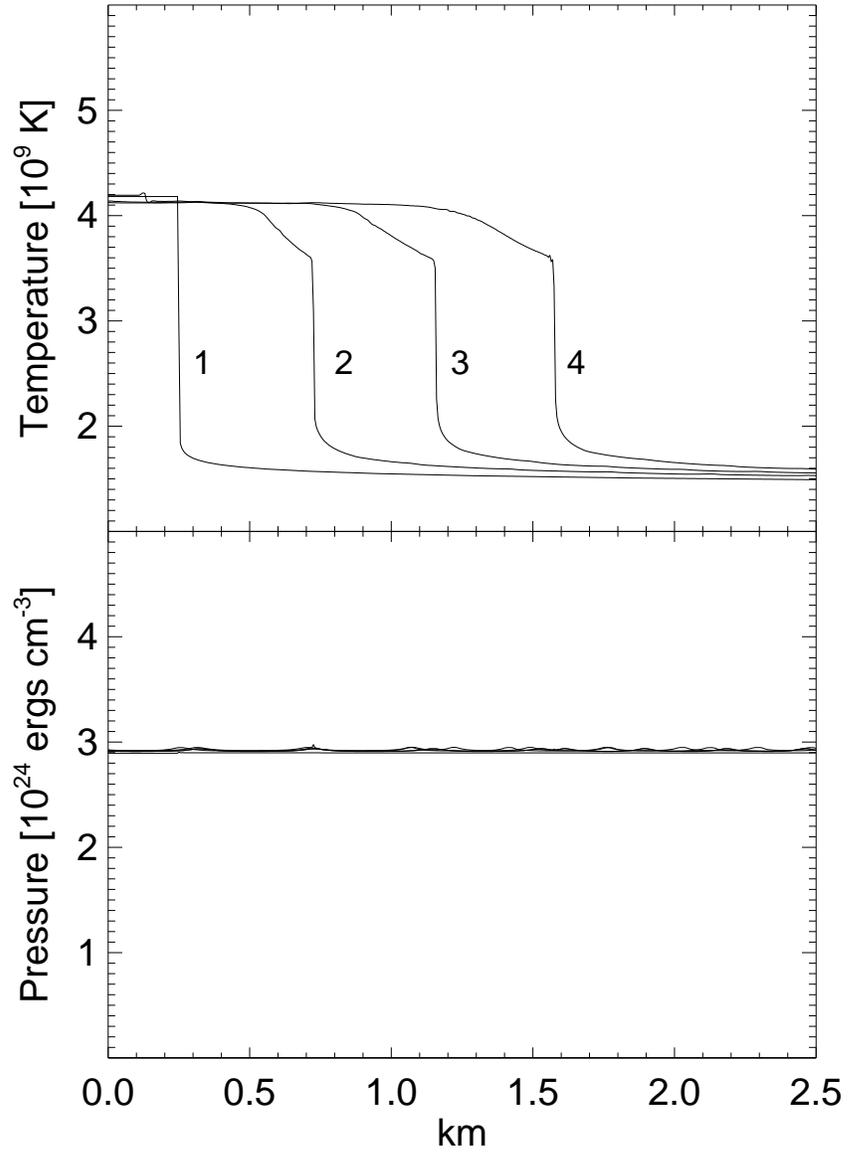}
\caption{Evolution of a spontaneous flame which does not
lead to a detonation. The matter has a density of  $ 2.3 \times 10^7 
\,\mbox{g cm}^{-3}$ with a carbon mass fraction of $ \onehalf$ and 
turbulent velocity is $0.2 \times 10^8 \mbox{cm s}^{-1}$. The 
numbers indicate subsequent times of the evolution: (1) 0 s, (2)
0.01 s, (3) 0.015 s and (4) 0.02 s.}
\end{figure*}

In Figure \ref{noddt2350} we see how a spontaneous wave
is initiated (at a density of $ 2.3 \times 10^7 \mbox{g cm}^{-3}$,
a fuel-composition of $X(^{12}\mbox{C}) = X(^{16}\mbox{O}) = \onehalf$ and
a turbulent velocity of $ u'(L) = 0.2\times 10^8 \mbox{cm s}^{-1}$)
and propagates along the temperature gradient. However, after an
elapsed time of $ 0.02 \mbox{s} = \tau_{\rm h}$ it does not pile up to a
detonation. Instead it moves with a constant velocity at constant
pressure.

\begin{figure*}[ht]
\figurenum{4}
\label{ddt2350}
\plotone{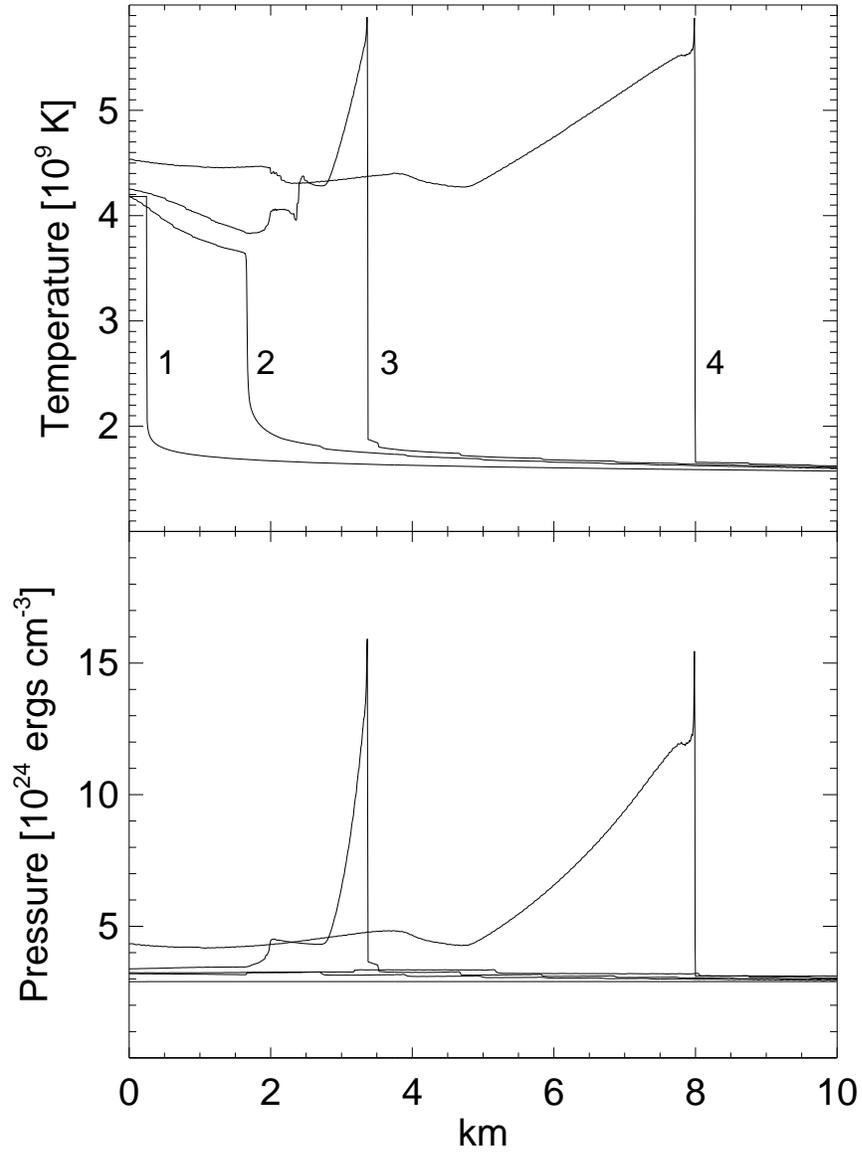}
\caption{Evolution to a detonation at a density of $ 2.3 \times 10^7 
\,\mbox{g cm}^{-3}$, a carbon mass fraction of $ \onehalf$ and a
turbulent velocity of $0.9 \times 10^8 \mbox{cm s}^{-1}$. The 
numbers indicate subsequent times of the evolution: (1) 0 s, (2)
0.0024 s, (3) 0.0028 s and (4) 0.0032 s.}
\end{figure*}

On the other hand, Figure \ref{ddt2350} shows the evolution from an
initial isobaric state to a detonation. The initial conditions
correspond to a turbulent velocity of $ u'(L) = 0.9 \times 10^8
\,\mbox{ cm s}^{-1}$ at a density of $ 2.3 \times 10^7 \,\mbox{ g
cm}^{-3}$ and a carbon mass fraction of $\onehalf$.
The initial flame profile immediately develops a spontaneous
burning front propagating along the spatial gradient of induction
timescales. The wave gradually accelerates and at a certain 
velocity the burning mode switches to a detonation.

Our results are summarized in Table \ref{tbl-1}, where we
give the turbulent velocity $ u'(L)$
necessary for the emergence of a detonation
out of certain initial conditions. The latter
are parametrized by the density and the nuclear composition
of the unburned matter. The values given in Table \ref{tbl-1}
are all obtained from numerical simulations performed in 
planar geometry. In spherical geometry -- due to spherical
damping -- these limits in terms of $ u'$ become higher. 

We find that even under assumptions in favour for a DDT 
the velocities are always 
higher than those expected in type Ia supernovae. 
This observation suggests that
a DDT could in principle only happen in the presence of stronger
turbulent velocity fluctuations than the 'typical' maximum value of 
$ 10^7 \,\mbox{cm s}^{-1}$, which results for Rayleigh-Taylor-driven turbulence. 
This conclusion holds for equal fractions of C and O and also for
mixtures enriched in carbon, although in the latter case the
constraint is weakened.

\section{Conclusions}

As a main result of our modeling we conclude that a transition from deflagration
to detonation triggered by turbulent mixing seems 
to be a possibility only if significantly more kinetic energy is stored in turbulence.
Our thresholds for the minimum values  
necessary for a DDT are all larger than the expected maximum r.m.s. turbulent
velocities obtained from numerical simulations of the explosion.
 Since we started from the most optimistic assumptions
(ignoring strong fluctuations or geometrical effects such as
spherical damping) we conclude that in more realistic models
these limits would become even higher. On the other hand, going to higher 
turbulent velocity fluctuations would then cause even stronger
fluctuations in temperature and fuel composition leading to 
a situation where a transition to detonation would become
practically impossible, as argued by Niemeyer (1999).
Thus it turns out that a DDT based on the Zeldovich gradient mechanism
must be regarded as a very unlikely scenario in type Ia supernovae.

\medskip
The authors thank M. Reinecke for providing us with the latest version
of PROMETHEUS.
This work was supported by the Deutsche Forschungsgemeinschaft and
the Deutscher Akademischer Austauschdienst.
AML also kindly acknowledges partial support by the NSF grants
INT-9726315 and AST-3731569.

%
%

\begin{deluxetable}{cccc}
\footnotesize
\tablecaption{Limiting threshold for the turbulent velocity
$ u'(L)$ at given density and fuel composition.
\label{tbl-1}}
\tablewidth{0pt}
\tablehead{
\colhead{$u'(L) $} 
& \colhead{$\rho $}
& \colhead{$ X(^{12}\mbox{C})$}
& \colhead{$ X(^{16}\mbox{O})$} \nl
\colhead{$[\mbox{cm s}^{-1}]$}
& \colhead{$[\times 10^7 \,\mbox{g cm}^{-3}]$}
& \colhead{}
& \colhead{}
}
\startdata
 $>0.5 \times 10^8$ &2.3  &0.5  &0.5  \nl
 $>0.6 \times10^8$ &1.3  &0.5  &0.5  \nl
 $>0.8 \times 10^8$ &0.8  &0.5  &0.5  \\[.2 cm]
 $>0.25 \times 10^8$ &2.3  &0.75  &0.25  \nl
 $>0.3 \times 10^8$ &1.3  &0.75  &0.25  \nl
 $>0.4 \times 10^8$ &0.8  &0.75  &0.25 \\[.2 cm]
 $>0.9 \times 10^8$ &2.3  &0.25  &0.75  \nl
 $>10^8$ &1.3  &0.25  &0.75  \nl
 $>10^8$ &0.8  &0.25  &0.75 \nl
\enddata
\end{deluxetable}

%
%

\pagebreak

\pagebreak

\pagebreak

%
%

\end{document}